\documentclass{article}

\usepackage{arxiv}
\usepackage[utf8]{inputenc} 
\usepackage[T1]{fontenc}    
\usepackage{hyperref}       
\usepackage{booktabs}       
\usepackage{amsfonts}       
\usepackage{nicefrac}       
\usepackage{textcomp,gensymb} 
\usepackage{siunitx}        
\usepackage{graphicx}
\usepackage{subcaption}     
\usepackage{mathtools}
\usepackage{indentfirst}
\usepackage{amsmath}

\newcommand{\eqdef}{\stackrel{\rm def}{=}} 
\DeclarePairedDelimiter\abs{\lvert}{\rvert}%
\DeclarePairedDelimiter\norm{\lVert}{\rVert}%

\makeatletter
\let\oldabs\abs
\def\abs{\@ifstar{\oldabs}{\oldabs*}}
\let\oldnorm\norm
\def\norm{\@ifstar{\oldnorm}{\oldnorm*}}
\makeatother

\title{Simulation and optimal control of heating and cooling systems: a case study of a commercial building}

\author{
  Phillipe R.~Sampaio \\
  Veolia Research and Innovation\\
  Maisons-Laffitte, France \\
  \texttt{sampaio.phillipe@gmail.com} \\
  \And
  Raphael Salvazet \\
  Veolia Research and Innovation\\
  Limay, France \\
  \texttt{raphael.salvazet@veolia.com} \\
  \And
   Pierre Mandel \\
  Veolia Eau d'Ile-de-France \\
  Nanterre, France \\
  \texttt{pierre.mandel@veolia.com} \\
  \And
  Gwénaëlle Becker \\
  Veolia Research and Innovation\\
  Limay, France \\
  \texttt{gwenaelle.becker@veolia.com} \\
  \And
  Damien Chenu \\
  Veolia Research and Innovation\\
  Limay, France \\
  \texttt{damien.chenu@veolia.com} \\
}

\begin{document}
\maketitle

\begin{abstract}
In this paper, an energy conservation measure that optimizes the planning of heating and cooling systems for tertiary sector buildings is proposed. It consists of a model-based predictive control approach that employs a grey-box model built from the building data history and from weather condition data that predicts the building heat load and indoor temperature. This model is then used by heating and cooling optimization strategies that aim at reducing the total energy consumption of the building in the next day while satisfying the desired indoor thermal comfort constraint. The proposed optimization strategies do not modify the regulation mode in place; rather, they send optimized set-points to the building management system in order to reduce the energy consumption. We applied our approach in a case study of a commercial building during heating and cooling seasons and we show that it was able to yield up to 12\% of energy savings while having a mean power forecast error of 8\%.
\end{abstract}


\section{Introduction}

In this work, we propose an energy conservation measure (ECM) that optimizes the planning of heating and cooling systems for tertiary sector buildings (e.g. offices, schools, hospitals). The proposed optimization strategies are part of the tool BatIntel developed by Veolia Research and Innovation (VERI) and they apply for buildings with intermittent heating or cooling management. In general, the comfort set-point temperature in a building must be maintained only when the building is occupied, in which case the system is said to be in comfort mode. In the remaining time, heating and cooling systems are in setback mode. The building operators take into account a time margin when switching from setback mode to comfort mode in order to ensure the thermal comfort for the arrival of the first occupants in the building. The switch is done at a fixed time following a schedule which is not always adjusted by the building operators. This simple and conservative way to manage the building is not efficient because the thermal needs of the building differ from one day to another and depend upon many variables, such as the weather conditions and distinct occupancy and ventilation operation schedule profiles for each day.

The objective of the BatIntel optimization strategies is to achieve the maximum energy savings potential for a given building while satisfying the desired indoor thermal comfort constraint. These optimization strategies do not modify the regulation mode in place; rather, they send optimized set-points to the building management system (BMS) in order to reduce the energy consumption. These strategies rely on a building model that predicts the heat load profile and indoor temperature for the next day. This model is a grey-box R6C2 model derived from the one proposed in \cite{Berthou2014} and it is built from the building data history and from weather condition data, as explained in due course. The choice of the R6C2 model as the building model comes from the fact that it compares favorably with others RC models as well as other physical and statistical models discussed in \cite{Berthou2014}. Although we employ a R6C2 model, our calibration procedure as well as the sunlight modelling differs substantially from those presented in \cite{Berthou2014}.

A full review of supervisory and optimal control techniques for building HVAC systems is done in \cite{Wang08} and in \cite{Foucquier2013}. In particular, model-based predictive control (MPC) has been proved to be a successful method for building climate control and for obtaining energy savings \cite{Oldewurtel2012, Ma2012}. Its greatest strengths relies on the fact that it takes into account not only the current timeslot to be optimized but all subsequent ones within a time interval while satisfying a set of constraints. 

Our approach for building thermal comfort belongs to the MPC category and builds on the results of \cite{Berthou2014} and \cite{Oldewurtel2012} for calculating optimal solutions for the control variables. It makes use of continuous black-box optimization algorithms rather than gradient-based algorithms for solving the underlying optimization problems. The computational time needed to obtain a good solution is thus quite low when compared to other approaches containing both discrete and continuous variables in the formulation and where more complex techniques such as mixed-integer nonlinear programming are applied. The proposed approach is easily applicable to both heating and cooling systems and requires only a simple model of the solar radiation to work well.

\textbf{Organization}. The outline of the paper is as follows. In Section 2, we first present the building model and we discuss about its calibration and the prediction accuracy criteria considered. In Section 3, we introduce the cooling and heating optimization strategies as well as the mathematical formulation that describes the optimization problem involved. In addition, we explain how we evaluate the potential energy savings. We detail the case study in the experimental results for both the cooling and heating cases in Section 4. Conclusions are given in Section 5.

\section{Building modelling}
\label{sec:buildingmodelling}

\subsection{Description of the model}
\label{subsec:modeldescription}

\indent The thermal modeling of the building is based on the equivalence between the thermal and electrical networks. This electrical-thermal analogy is shown in Table~\ref{table:electric-thermal}. We employ a resistance-capacity (RC) model composed of six thermal resistances representing the different forms of thermal losses (conduction and convection) and two capacities representing the thermal storage in the building.

\begin{table}[htb]
\caption{Electric-thermal analogy}
   \centering
   \begin{tabular}{l  l}
     \hline
     \textbf{Thermal}            & \textbf{Electric equivalence} \\
     \hline
     Thermal flow ($W$)          & Electric current ($A$) \\
     Temperature ($\degree$C)    & Potential ($V$)  \\
     Thermal resistance ($K/W$)  & Electric resistance  ($\si{\ohm}$)  \\
     Thermal capacitance ($J/K$) & Electric capacitance (Farad)  \\
     \hline
   \end{tabular}
\label{table:electric-thermal}
\end{table}

The RC model, which we will refer to as the BatIntel model in this paper, is a data-driven deterministic grey-box model whose parameters -- resistances and capacities -- are calibrated by an optimization algorithm. After the calibration is done, this model is used to predict the building indoor temperature and thermal power. The inputs and outputs of the BatIntel model are described in detail in Table~\ref{table:modelinout} while its RC parameters are shown in Table~\ref{table:modelparams}.

\begin{table}[htbp]
\caption{Input data used for calibrating the model and its outputs.}
   \centering
   \begin{tabular}{l  l  c  l}
     \hline
             & \textbf{Name}  & \textbf{Unit} \\
     \hline
     Outputs         & Forecast indoor temperature & $\degree$C & \\
                     & Forecast thermal power & $kW$ & \\
     Dynamic inputs  & Historical thermal power data & $kW$ &\\
                     & Historical indoor temperature data & $\degree$C & \\
                     & Historical external temperature data & $\degree$C & \\
                     & Historical solar irradiance data & $W/m^2$ & \\
     Static inputs   & Occupancy profile & 0 or 1 & \\
                     & Day set-point indoor temperature & $\degree$C & \\
                     & Night set-point indoor temperature & $\degree$C & \\
                     & Start/end time of day set-point indoor temperature & HH:mm & \\
                     & Ventilation operation schedule & 0 or 1 & \\
                     & Minimum and maximum power of the thermal production system & $kW$ & \\
                     & Total volume of the building & $m^3$ & \\
                     & Efficiency of heating or cooling machines & $[0,1]$ & \\
                     & Lower heating value (LHV) & $kWh/m^3$ & \\
     \hline
   \end{tabular}
\label{table:modelinout}
\end{table}

\begin{table}[htbp]
\caption{Description of the BatIntel model parameters.}
   \centering
   \begin{tabular}{l  l  l}
     \hline
     \textbf{Name} & \textbf{Description}                    & \textbf{Unit} \\
     \hline
     $R_i$    & Interior convective resistance                  & $K/W$ \\
     $R_m$    & Outdoor wall conductive resistance              & $K/W$ \\
     $R_s$    & Indoor wall conductive resistance               & $K/W$ \\
     $R_f$    & Infiltration and glazing equivalent resistance  & $K/W$ \\
     $C_i$    & Internal air capacitance                        & $J/W$ \\
     $C_m$    & Wall capacitance                                & $J/W$ \\
     $R_v$    & Mechanical ventilation equivalent resistance    & $K/W$ \\
     $R_e$    & External convective resistance                  & $K/W$ \\
     $g$      & Maximum heat gain due to occupancy              & $W$   \\
     $\alpha$ & Coefficient for solar irradiance gains          & $cm^2/s$ \\
     $a$      & Radiative ratio of internal gains               &  -     \\
              & (the convective ratio equals $1-a$)             &       \\
     \hline
   \end{tabular}
\label{table:modelparams}
\end{table}

\begin{figure}[ht]
     \centering
     \includegraphics[scale=0.42]{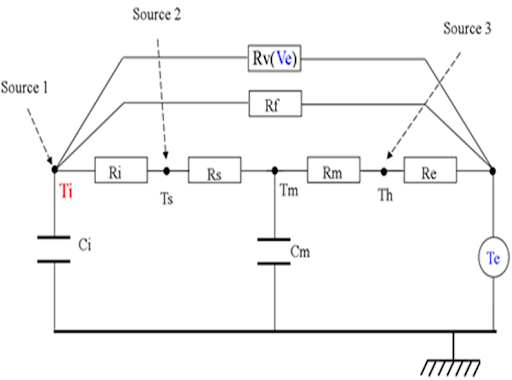}
     \caption{Electric circuit representation of the thermal network with 6 resistors and 2 capacitors.}
     \label{fig:r6c2_diagram}
\end{figure}

The corresponding electric circuit is illustrated in Figure~\ref{fig:r6c2_diagram}. The Source 1 represents the internal convective gain ($W$) and it is defined by 
\begin{equation}\label{eq:source1}
\textrm{Source 1} \eqdef \left\{
  \begin{array}{l}
    (1-a) \cdot g \cdot OCC(t) + P(t), \\
    P_{\min} \leq P(t) \leq P_{\max},
  \end{array}
\right.
\end{equation}
where $a$ is the radiative ratio of internal gains, $g$ is the maximum heat gain due to occupancy, $P(t)$ is the electric power at time $t$, $P_{\min}$ is the minimum electric power, $P_{\max}$ is the maximum electric power and $OCC(t)$ is the occupancy ratio that equals 1 when the building is occupied or 0 otherwise. The Source 2 denotes the internal radiative gain ($W$), being defined by
\begin{equation}\label{eq:source2}
\textrm{Source 2} \eqdef a \cdot g \cdot OCC(t) + \Phi_{\textrm{int}}(t).
\end{equation}
where $\Phi_{\textrm{int}}(t)$ is the solar flux on indoor walls. Finally, the Source 3 indicates the external radiative gain ($W$) that is defined by
\begin{equation}\label{eq:source3}
\textrm{Source 3} \eqdef \Phi_{\textrm{ext}}(t),
\end{equation}
where $\Phi_{\textrm{ext}}(t)$ is the solar flux on outdoor walls.
The $T_i$ node represents the mean indoor temperature while the $T_m$ node denotes the temperature of the center of the wall of the building. The two other nodes are the outdoor surface wall node, $T_h$, and the indoor surface wall, $T_s$.

The BatIntel model is then expressed by the following differential-algebraic equations (DAEs), which are solved by a Runge-Kutta method:
\begin{subequations}\label{eq:daemodel}
\begin{align}
\frac{T_h(t)(R_m + R_e)}{R_m R_e} &= \frac{T_m(t)}{R_m} + \frac{T_e(t)}{R_e} + \alpha \cdot \Phi_{\textrm{ext}}(t) \\
\frac{T_s(t)(R_i + R_s)}{R_i R_s} &= \frac{T_i(t)}{R_i} + \frac{T_m(t)}{R_s} + a \cdot g \cdot OCC(t) \\
\overline{P}(t) &= C_i \cdot \frac{(T_{\textrm{set-point}}-T_i(t))}{dt} + \frac{(T_i(t)-T_s(t))}{R_i} + \frac{(T_i(t)-T_e(t))}{R_f} - \\ 
& (1-a) \cdot g \cdot OCC(t) + R_v \cdot VEN(t) \cdot (T_i(t)-T_e(t))  \\
P(t) &= \max (\min (\overline{P}(t), P_{\max}), P_{\min}) \label{eq:subeqpower} \\
C_i \frac{dT_i(t)}{dt} &= \frac{(T_s(t)-T_i(t))}{R_i} + \frac{(T_e(t)-T_i(t))}{R_v} + \frac{(T_e(t)-T_i(t))}{R_f} + (1-a) \cdot g \cdot OCC(t) + P(t) \label{eq:subeqtemp} \\
C_m \frac{dT_m(t)}{dt} &= \frac{(T_h(t)-T_m(t))}{R_m} + \frac{(T_s(t)-T_m(t))}{R_s}
\end{align}
\end{subequations}
where $T_{\textrm{set-point}}$ is a set-point temperature defined a priori that varies from day time to night time, $VENT(t)$ represents the ventilation airflow schedule of the building with values equal to 0 or 1 and $P(t)$ is the electric power estimation which is bounded by thresholds that depend on the building system specifications.

The solar irradiance is expressed in the model in Watts per square meter ($W/m^2$) and it quantifies the power per unit of horizontal area of solar radiation provided as input data. Our sunlight approach is therefore less complex than the one presented in \cite{Berthou2014}, where they make use of the angle of solar radiation and the solar orientation as well as solar masks to calculate the radiation data for each building. Nevertheless, our approach has the advantage that it allows to work with easily available meteorological data. In our case study, we set $\Phi_{\textrm{int}}(t)=0$ and only consider the external radiative gain $\Phi_{\textrm{ext}}(t)$.

It is important to note that the model used here can represent only one thermal zone, thus assuming that all areas of the building are cooled or heated in the same way. One of the next steps for improving the current research work is to extend this model to handle multiple thermal zones and benefit from the optimization framework already built.

\subsection{Model calibration}
\label{subsec:modelcalibration}

\begin{figure}[ht]
     \centering
     \includegraphics[scale=0.6]{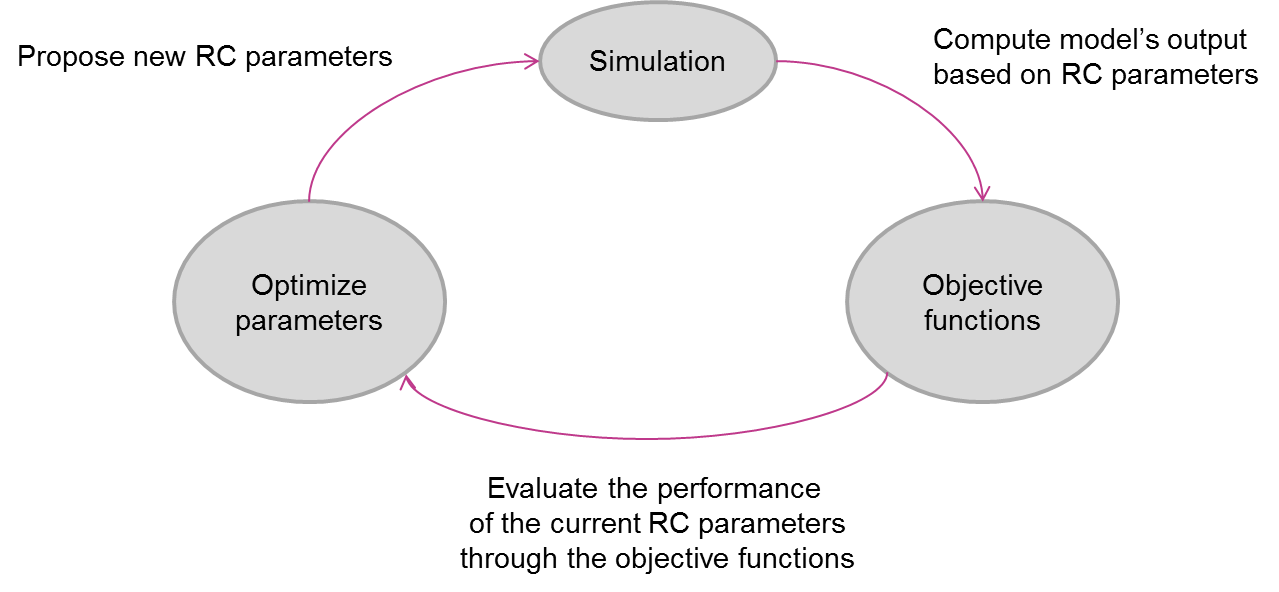}
     \caption{Calibration procedure of the BatIntel model.}
     \label{fig:calibflow}
\end{figure}

The calibration of the model consists in identifying the values of the internal parameters described in Table~\ref{table:modelparams} that minimize the difference between the forecast values and the actual values of power and indoor temperature using a historical data set of four weeks. The number of weeks was chosen after verifying that no significant accuracy improvement came from increasing the number of days and that the optimal value found approached four weeks in our experiments. Figure \ref{fig:calibflow} illustrates the calibration procedure through diagrams where a new set of RC parameters is proposed every iteration until the convergence criteria are satisfied.

The optimization variables in the calibration procedure, also known as control variables in the field of optimal control, are the parameters described in Table~\ref{table:modelparams}, whereas the objective functions to be minimized are:
\begin{itemize}
    \item $f_1$: the difference between forecast and actual indoor temperature values;
    \item $f_2$: the difference between forecast and actual thermal power values.
\end{itemize}

Due to the fact that both the power and the indoor temperature indicators ($f_1$ and $f_2$) are taken into account in the evaluation of the quality of the model, the calibration procedure can be seen as a problem of multiobjective optimization. In this type of problem, it is rare to find a single solution that optimizes each of the objectives simultaneously. Instead, we seek to identify a list of solutions, called Pareto solutions, that can not be improved without one of the objective functions being degraded. The set of all of these solutions is called Pareto frontier.

We define the first objective function with the \textit{Mean Absolute Percentage Error} (MAPE) measure for evaluating the prediction accuracy for the indoor temperature. This measure represents the relative error in indoor temperature and is defined as follows:
\begin{eqnarray}\label{eq:temperror}
f_1(\sigma) \eqdef \frac{100}{n} \sum_{t=1}^{n} \abs{\frac{\overline{T}_t(\sigma) - T_t}{T_t}},
\end{eqnarray}
where $\sigma$ is the vector of control variables, $T_t$ is the actual value at instant $t$, $\overline{T}_t(\sigma)$ is the forecast value at instant $t$ for a given $\sigma$ and $n$ is the total number of fitted points. Note that $\overline{T}_t(\sigma)$ is calculated by Equation \eqref{eq:subeqtemp} and that all forecast values in this paper are overscored while the actual values are not. The only exception is in the definition of the DAEs in Equations \eqref{eq:daemodel}, in which case, since there are no actual power and indoor temperature values involved therein, we do not overscore the forecast variables for the sake of simplicity.

The second objective function represents the absolute error in power and is defined by the median of the absolute deviation between the forecast and the actual values in power over time:
\begin{eqnarray}\label{eq:powerabserror}
f_2(\sigma) \eqdef median\{\abs{\overline{P}_t(\sigma) - P_t}\},
\end{eqnarray}
where $P_t$ is the actual value at instant $t$ and $\overline{P}_t(\sigma)$ is the forecast value at instant $t$ for a given $\sigma$ calculated by Equation \eqref{eq:subeqpower}. We chose the median among the statistical measures for calculating the error in power because it is less sensitive to outliers than the mean and because it mitigates the impact of noise present in signals, which is the case with the power data in our case study. Furthermore, we do not divide the result of $\overline{P}_t(\sigma) - P_t$ by $P_t$ as it is done in \eqref{eq:temperror}, because the power values can be close or equal to zero. For this reason, the normalization is done afterwards using a min-max approach as explained in the next subsection.

The multiobjective optimization problem of the calibration process is then expressed as
\begin{equation}\label{eq:calibopt}
\begin{array}{rc}
\text{minimize}_{\sigma} \quad & (f_1(\sigma), f_2(\sigma)) \\
\text{subject to}        \quad & l \leq \sigma \leq u,
\end{array}
\end{equation}
where $l$ and $u$ are lower bounds and upper bounds for each entry of $\sigma$, respectively.
This problem is actually a dynamic multiobjective optimization problem involving DAEs, which we address by coupling a DAE solver with a multiobjective optimization algorithm. Our approach differs from the "all-at-once" method where the discretized DAEs are incorporated into the optimization formulation (see Chapters 9 and 10 in \cite{Biegler2010} for more details about both approaches). Moreover, we do not apply a derivative-based nonlinear programming (NLP) method to solve the optimization problem, but rather a black-box optimization method. By doing this, the BatIntel model and the optimization problem are decoupled, which allows more flexibility when one of them must be changed in the future. Besides, this approach avoids the difficulties related to the derivatives of the functions such as their existence and smoothness, which are often required in the convergence theory of many optimization algorithms, and it also reduces the chances of getting stuck on local optima in early iterations as most of the derivative-based NLP solvers are designed to find local optima only. 

In order to solve \eqref{eq:calibopt}, the Non-dominated Sorting Genetic Algorithm II (NSGA-II) \cite{Deb2002} has been applied. A genetic algorithm is a meta-heuristic inspired by the process of natural selection and relies on operations of mutation, crossover and selection of the iterates to produce better solutions. In addition to these three operations, the NSGA-II has two multi-objective operators that allows to build the Pareto frontier:
\begin{itemize}
    \item Fast non-dominated sorting: the population is sorted and partitioned into different non-domination levels;
    \item Crowding distance: a ranking mechanism used in the selection phase to preserve solution diversity.
\end{itemize}
The choice of NSGA-II in the the calibration procedure is twofold: (1) it compares favorably with other multi-objective algorithms on many academic and engineering applications \cite{Deb2002}, having been applied in a wide range of real-life applications, and (2) it is easily applicable when the functions are the result of a simulation process.

Since we have to choose one single parameter solution in the end to proceed to the optimization strategy, we take the best compromise solution between the two objectives. To achieve this goal, we implemented the TOPSIS (Technique for Order of Preference by Similarity to Ideal Solution) \cite{Hwang47} multicriteria method. This method chooses the solution that is the closest to the ideal positive solution and the farthest from the ideal negative solution, after normalizing the scores for each objective function and after calculating the geometric distances between Pareto solutions and ideal solutions.

\subsection{Prediction accuracy criteria}
\label{subsec:predacccriteria}

After the calibration is done and a Pareto solution $\hat{\sigma}$ has been selected, the following three accuracy criteria must be satisfied by the resulting model in order to use it in the BatIntel optimization strategy:
\begin{enumerate}
    \item Median absolute error in indoor temperature < $TEMP_{error}^{absolute}$,
    \item Relative error in indoor temperature $T_{\textrm{error}}$ < $TEMP_{error}^{relative}$,
    \item Relative error in power consumption $P_{\textrm{error}}$ < $POWER_{error}^{relative}$,
\end{enumerate}
where
\begin{equation}\label{eq:temppowererrors}
T_{\textrm{error}}(\hat{\sigma})\eqdef f_1(\hat{\sigma}), \quad P_{\textrm{error}}(\hat{\sigma}) \eqdef \frac{f_2(\hat{\sigma})}{P_{\textrm{max}}-P_{\textrm{min}}},
\end{equation}
$P_{\textrm{max}}$ and $P_{\textrm{min}}$ are the maximum and minimum actual power values, respectively, and $TEMP_{error}^{absolute}$, $TEMP_{error}^{relative}$ and $POWER_{error}^{relative}$ are constants whose values in our case study are 1$\degree$C, 5\% and 12\%, respectively. 

\section{The BatIntel optimization strategy}
\label{sec:optstrategy}

\subsection{General description}
\label{subsec:strategydesc}

The BatIntel optimization strategy aims at reducing the energy consumption (cooling and heating) of the building while satisfying the comfort requirement for the indoor temperature. The three decision variables (only two for cooling) involved and their description are detailed below:
\begin{enumerate}
    \item Day initial time: switch over from night set-point to comfort (day) set-point;
    \item Night set-point indoor temperature: set-point indoor temperature to be attained in the building during night time; 
    \item Night initial time: switch over from day set-point to night set-point.
\end{enumerate}

Depending on the local weather, the night set-point temperature might not be as important for the cooling strategy as it is for the heating strategy as it may not be necessary to cool the building during the night in order to attain the comfort temperature in the beginning of the day. We observed this event in the case study of the building in Brussels described in Section~\ref{sec:casestudy}  and decided not to consider the night set-point temperature as variable for the cooling strategy in that case. This kind of event occurs, for instance, when the external temperature during the night is below 20$\degree$C, thus being already below the comfort temperature required. Since there is no occupancy in the building before 7:00 AM, the indoor temperature will not increase and, therefore, there is no need for cooling the building during night time.

\subsection{Mathematical formulation of the problem}
\label{subsec:objfuncons}

We describe now the formulation of the associated optimization problem. The objective function consists in minimizing the total energy consumption predicted by the BatIntel model for the next day, which is calculated by the sum of power predictions at discrete times with 15-minute intervals. Since the sample interval is constant, it serves only as a scaling parameter for the sum of power predictions in the objective function; therefore, we can dispose of it as the optimization results will not be affected. As for the constraint related to the comfort temperature, we require the latter to be an upper bound for the indoor temperature in the cooling strategy and a lower bound in the heating strategy within a time interval ranging from $t_{beg}^{comf}$ to $t_{end}^{comf}$. The optimization problem for the cooling strategy is then formulated as
\begin{equation}\label{eq:optstrategy}
\begin{array}{rc}
\text{minimize}_{\theta} \quad & f(\theta) \\
\text{subject to}        \quad & g(\theta) \leq 0, \\
                         \quad & l \leq \theta \leq u,
\end{array}
\end{equation}
where 
\begin{equation}\label{eq:objfunstrat}
\begin{array}{c}
f(\theta) \eqdef \sum_{t}\overline{P}_t(\theta), \\
g(\theta) \eqdef \max_{t=t_{beg}^{comf},\ldots,t_{beg}^{comf}} \{ \overline{T}_t(\theta)-T_{\textrm{comf}} \}, \\
\end{array}
\end{equation}
$\theta$ is the vector containing the three decision variables described in Section \ref{subsec:strategydesc}, $l$ and $u$ are lower bounds and upper bounds on each entry of $\theta$, respectively, $\overline{P}_t(\theta)$ and $\overline{T}_t(\theta)$ are the power and the indoor temperature at instant $t$ predicted by the BatIntel model, respectively, and $T_{\textrm{comf}}$ is the comfort set-point temperature defined a priori. The only difference of this formulation from the one used in the heating strategy is the inequality sign in the constraint related to $g(\theta)$ in \eqref{eq:optstrategy}, in which case it is flipped.

In \eqref{eq:optstrategy}, it is demanded that the forecast indoor temperatures, $\overline{T}_t(\theta)$, be lower than or equal to the comfort set-point temperature, $T_{\textrm{comf}}$, within the time interval from $t_{beg}^{comf}$ to $t_{end}^{comf}$ . Rather than having one constraint per instant $t$ to express this comfort requirement, we consider the maximum value of $\overline{T}_t(\theta)-T_{\textrm{comf}}$ from all $t$ and we check if this value is negative, zero or positive. If it is strictly positive, it means that $\overline{T}_t(\theta) > T_{\textrm{comf}}$ for at least one instant $t$, which implies that the comfort requirement is not satisfied in the predefined time interval. Otherwise, if the constraint in \eqref{eq:optstrategy} is satisfied for the maximum value of $\overline{T}_t(\theta)-T_{\textrm{comf}}$, then $\overline{T}_t(\theta)\leq T_{\textrm{comf}}$ for every instant $t$. By using the $\max$ function for modelling the comfort requirement, we considerably reduce the number of constraints in the problem. For instance, assuming that $t_{beg}^{comf}$ = 8AM and $t_{end}^{comf}$ = 8PM, as in our case study, rather than having 180 (the number of 15-minute intervals from 8AM to 8PM) constraints we end up with one single constraint, which simplifies the problem. This is of great importance as a large number of constraints usually increases the computational time and makes the problem more difficult to be solved by most of the optimization algorithms. On the other hand, the fact that the $\max$ function is not differentiable could make the problem more difficult to be solved by optimization algorithms that require smoothness of the functions. However, we use an optimization algorithm that neither requires the derivatives of the functions nor their smoothness as the functions are considered black boxes. It belongs to the category of black-box optimization methods called "direct search". 

Direct-search methods are iterative methods that do not require the analytic expressions of the functions or derivatives, but only the function values. These methods are based on comparing the values of the functions at several points in a mesh at each iteration. If the algorithm finds a new point in the mesh that improves the objective function (total energy consumed in the day, for example) with respect to the current point, the new point becomes the current point in the next iteration of the algorithm.
 
Among the existing techniques in the class of direct-search methods, the mesh adaptive direct search methods (MADS) \cite{Audet2006} proved to be very efficient in practice and one of the appealing features in contrast to other derivative-free algorithms is the existence of a supporting convergence theory. A popular solver that implements this algorithmic approach and that was chosen in our experiments is NOMAD \cite{Nomad}. Other black-box optimization methods that make use of surrogate models to replace the black-box functions during the optimization process (see, for instance, \cite{Sampaio2015,Sampaio2016,Powell94c,Powell98b, LewisTorczon02,LucidiSciTseng2002}) are also commonly used in real-life applications and could be applied to solve the problem \eqref{eq:optstrategy}. A comparative study of the performance of different black-box algorithms on \eqref{eq:optstrategy} is left for future research work. 

Notice that the problem \eqref{eq:optstrategy} is also a dynamic optimization problem involving DAEs and that we solve it by coupling a DAE solver with a black-box optimization algorithm in a way similar to what we have done in the calibration process. In order to obtain the best available solution, we use Latin Hypercube Sampling (LHS) \cite{McKay79} as a multistart method for generating the starting points of many optimization runs. LHS is a statistical sampling approach widely used for design of experiments that allows to construct samples from a probability distribution while ensuring that the search space is well covered.

\subsection{Performance criteria}
\label{subsec:performancecriteria}

In order to evaluate the performance of the BatIntel optimization strategy, we first calculate the energy consumption reduction obtained with the optimization over each day of the test period and then we take the mean of all days over the entire test period. For this purpose, we consider the actual data of the energy consumed after the application of the optimized set-points proposed by BatIntel (i.e., actual optimized energy) and compare it to the energy that would be consumed in the absence of optimization on the same day when applying the typical set-points used in the building. However, one can not have the actual data of the two power curves over the same period since the application of the optimized set-points prevents the existence of the other. Therefore, the “non-optimized energy” is simulated by the BatIntel model (forecast non-optimized energy) and its value is used in the calculation of the savings.
Assuming that the power value is constant within the uniform time intervals $\Delta t$, the formula for estimating the energy consumption reduction, or simply, the energy savings ($ES$) for one single day is then defined as
\begin{equation}\label{eq:eg}
ES \eqdef \displaystyle\frac{\overline{E}^{\,\textrm{non-opt}} - E^{\,\textrm{opt}}}{\overline{E}^{\,\textrm{non-opt}}} = \displaystyle\frac{\Delta t\sum_{t=t_{beg}^{savings}}^{t_{end}^{savings}} \left(\overline{P}_t - P_t\right)}{\Delta t\sum_{t=t_{beg}^{savings}}^{t_{end}^{savings}} \overline{P}_t} = \displaystyle\frac{\sum_{t=t_{beg}^{savings}}^{t_{end}^{savings}} \left(\overline{P}_t - P_t\right)}{\sum_{t=t_{beg}^{savings}}^{t_{end}^{savings}} \overline{P}_t},
\end{equation}
where $\overline{E}^{\,\textrm{non-opt}}$ is the forecast non-optimized energy consumed during the day, which is the time integral of the forecast non-optimized power, and $E^{\,\textrm{opt}}$ is the actual optimized energy consumed during the day, being the time integral of the actual optimized power. The savings is calculated by taking the time interval ranging from $t_{beg}^{savings}$ to $t_{end}^{savings}$, whose values used in the experiments are $t_{beg}^{savings}=7PM$ and $t_{end}^{savings}=10PM$ for the cooling season and $t_{beg}^{savings}=4AM$ and $t_{end}^{savings}=20AM$ for the heating season. The reason for having a small time range for the cooling season in our case study is explained in due course.

\section{Case study}
\label{sec:casestudy}

\subsection{About the building and the data}
\label{subsec:thebuilding}

The BatIntel model is intended to medium- and high-energy buildings and it needs historical data of the building operation. Besides, it is a monozone model and thus it requires a control level of the heating and cooling systems on a global scale of the building. Learning of these requirements, we selected a building located in Brussels, hereafter denoted as the BatIntel building for the sake of confidentiality, that has significant heating and cooling consumption to be the our demonstrator building. It is also monitored, which allows to have the required data, and equipped with a Building Management System (BMS) that enables the automatic regulation of the heating and the cooling systems. The general characteristics of the BatIntel building are given in Table~\ref{table:batintelbuilding}.

\begin{table}[ht]
\caption{General characteristics of the BatIntel building.}
   \centering
   \begin{tabular}{l  l  l}
     \hline
     \textbf{General info}     & \textbf{Surface area} & Nearly 25,000m$^2$. \\
                      & \textbf{Number of floors} & Five floors (plus three parking lot basements).  \\
                      & \textbf{Construction year} & 2009. \\
                      & \textbf{Inertia} & Medium. \\
                      & \textbf{Occupancy} & Mostly between 7AM and 9PM from Monday to Friday. \\
                      & \textbf{Usage} & 81\% open space and rest rooms, 9\% restoration, 4\% sanitary, \\
                      &                & 2\% computer room and 4\% others. \\
    \hline
     \textbf{Technical systems} & \textbf{Heating production} & Three gas boilers: two $654kW$ boilers and one $327kW$ boiler.  \\
                      &  & Total maximum power: $1635kW$. \\
                      & \textbf{Cooling production} & Two $1MW$ refrigeration units. \\
                      & \textbf{Ventilation} & Eight air handling units. \\
     \hline
   \end{tabular}
\label{table:batintelbuilding}
\end{table}

In what follows, we present the numerical results of the BatIntel solutions for cooling and heating seasons applied in the BatIntel building in the summer of 2017 and in the winter of 2017-2018. The model calibrations were done with historical data of indoor temperature and thermal power of 28 days, at a 15-minute time interval. The measured and the forecast external temperature and solar irradiance data came from the Royal Meteorological Institute of Belgium (IRM) and were at an one-hour time interval, being then linearly interpolated to be at the same 15-minute time interval of the building measured data. Since the model was re-calibrated every week with new data in order to improve its prediction accuracy, we show here the results from only one calibration procedure done for heating and one for cooling. For both seasons, the time interval where the comfort-related constraint in the problem \eqref{eq:optstrategy} must be satisfied ranges from $t_{beg}^{comf}=8AM$ to $t_{end}^{comf}=8PM$. In the heating case, $T_{\textrm{comf}}=23\degree$C, and, in the cooling one, $T_{\textrm{comf}}=24\degree$C.

\subsection{Cooling season}
\label{subsec:coolingseason}

\subsubsection{Calibration results}
\label{subsec:coolingcalib}

\begin{figure}
    \centering
    \begin{subfigure}[t]{\textwidth}
        \centering
        \includegraphics[height=2in]{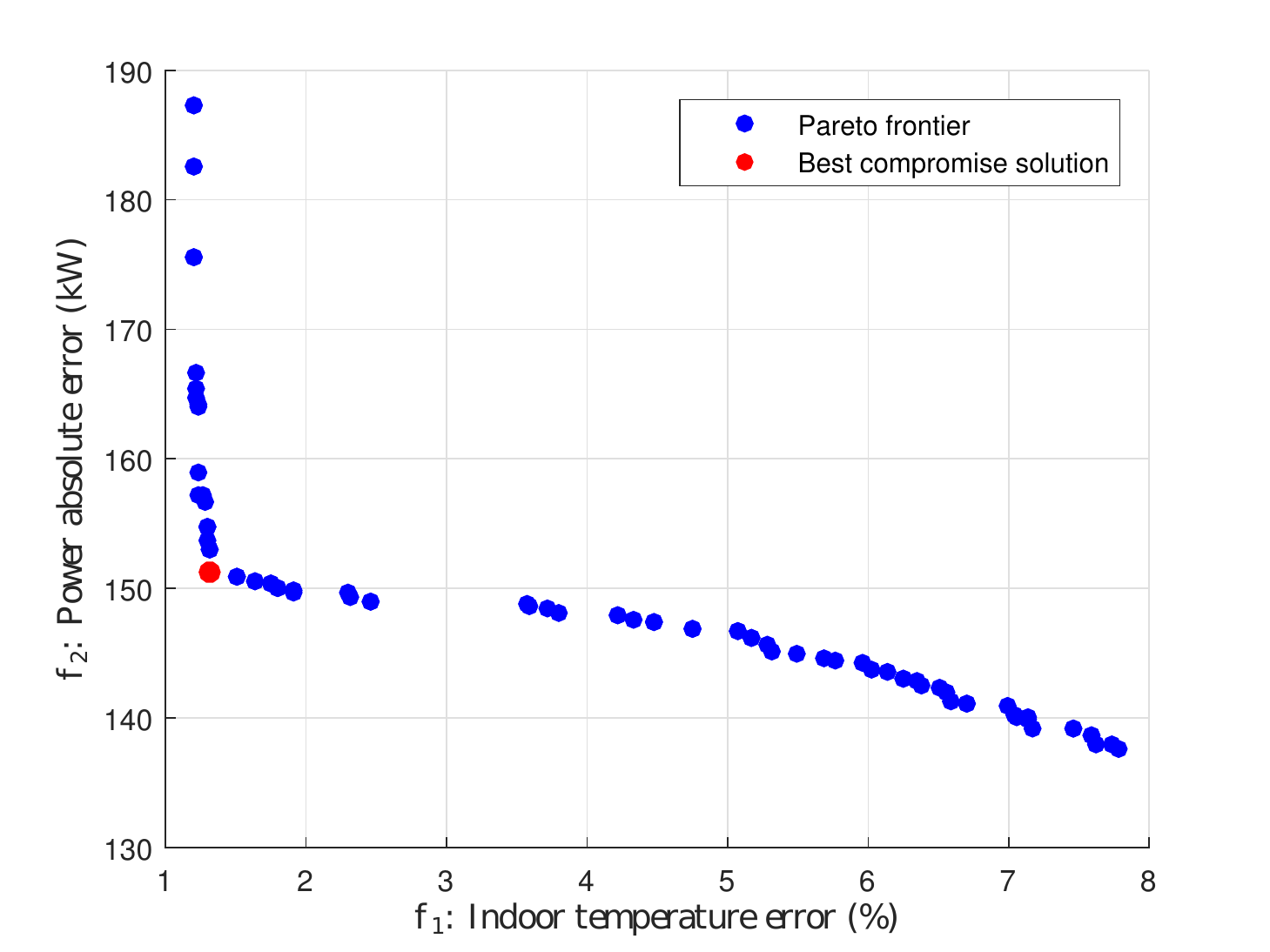}
        \caption{Pareto frontier obtained from the cooling season. In the vertical axis, the median absolute deviation in power ($kW$), and, in the horizontal axis, the relative error in indoor temperature.}
        \label{fig:paretofrontier_cooling}
    \end{subfigure}
    \begin{subfigure}[t]{0.5\textwidth}
        \centering
        \includegraphics[height=2in]{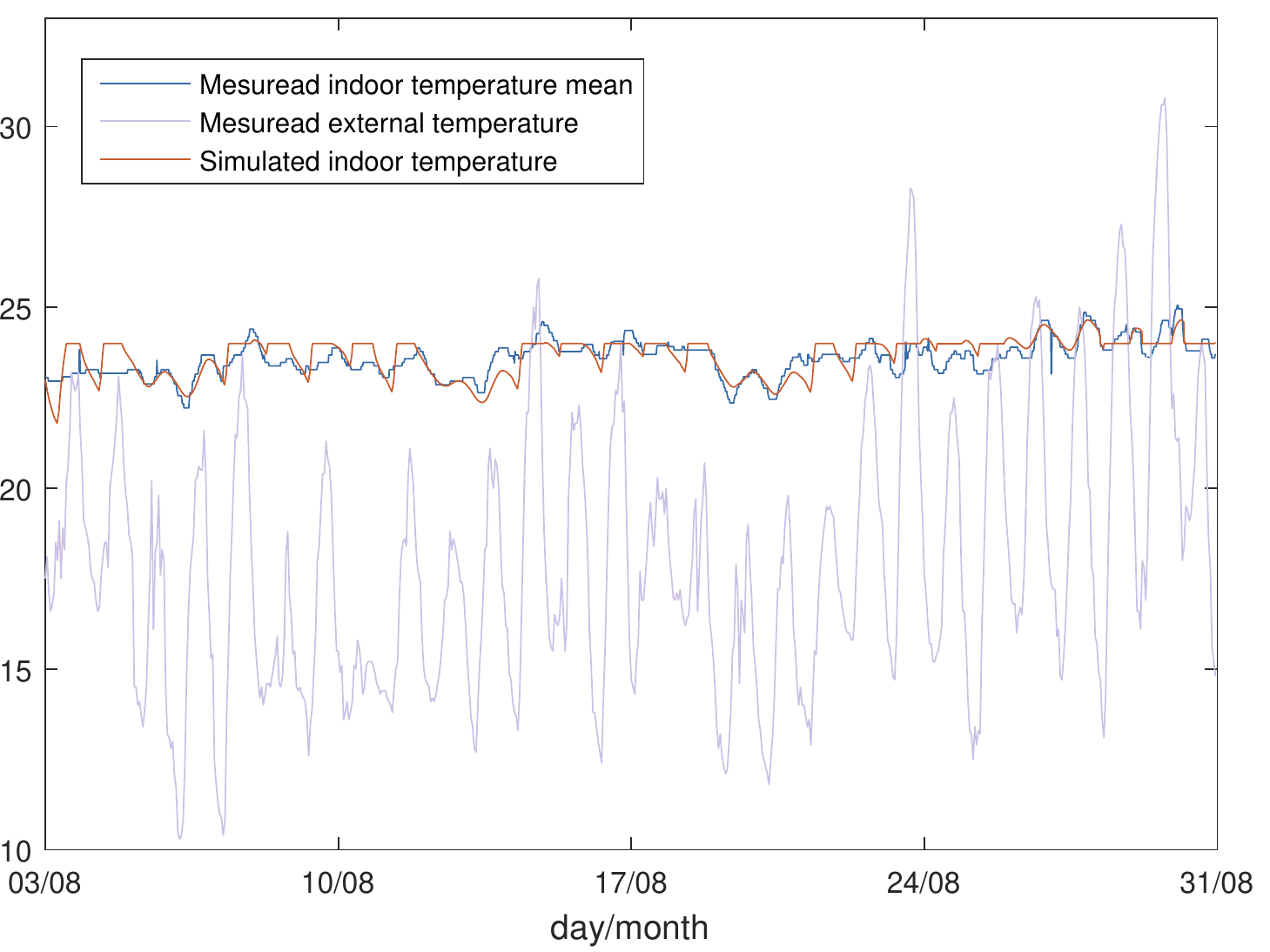}
         \caption{Temperature curves ($\degree$C) obtained with the best compromise solution from the Pareto frontier over the cooling season. In blue, actual indoor temperatures; in red, indoor temperatures predicted by the model; and, in violet, actual external temperatures.}
        \label{fig:tempplotscooling}
    \end{subfigure}%
    ~ 
    \begin{subfigure}[t]{0.5\textwidth}
       \centering
        \includegraphics[height=2in]{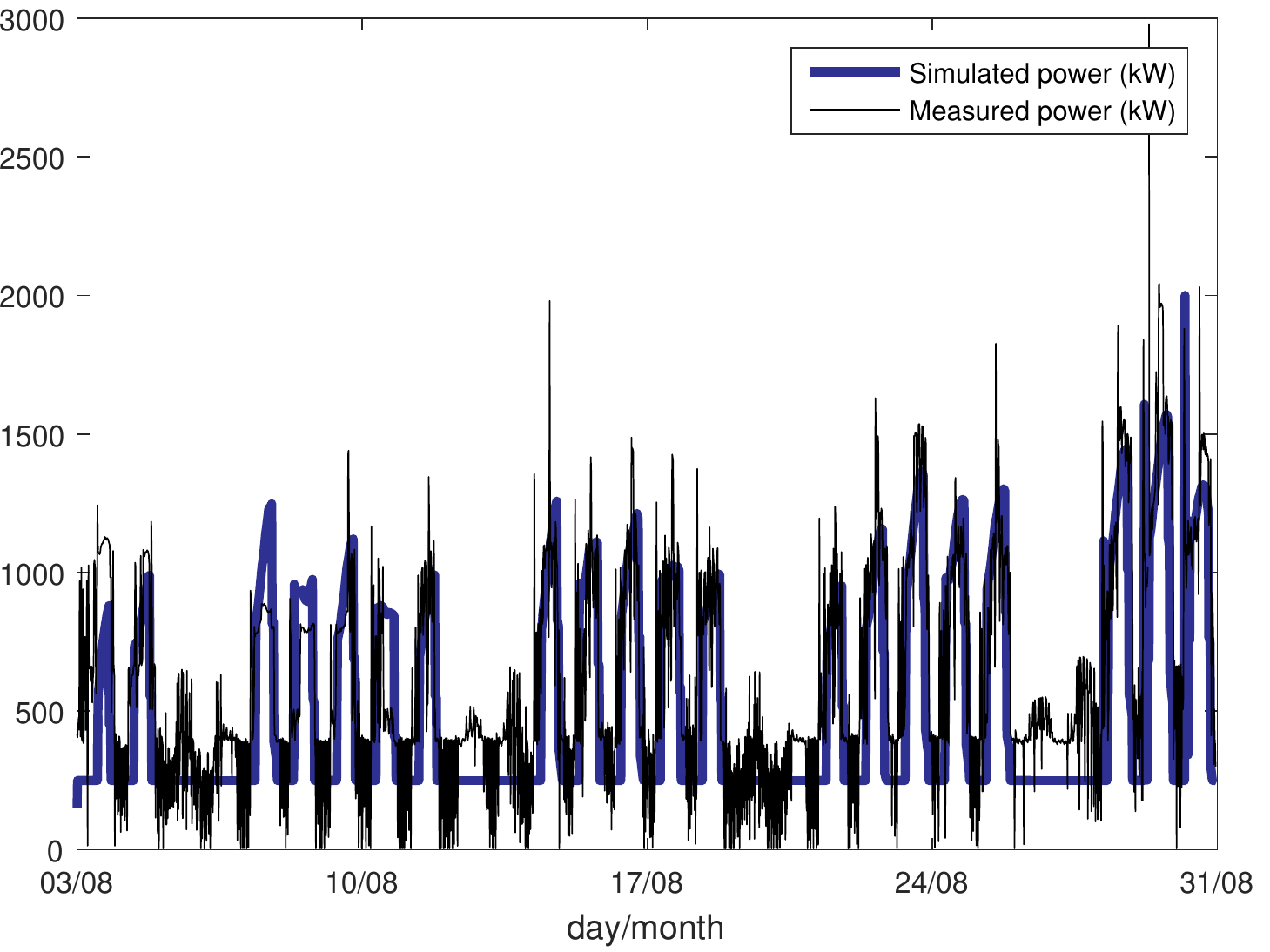}
        \caption{Power curves obtained with the best compromise solution from the Pareto frontier over the cooling season. In black, the actual power, and, in blue, the power predicted by the model.}
        \label{fig:powerplotscooling}
    \end{subfigure}
    \caption{Calibration results from the cooling season with data from 03/08/2017 to 30/08/2017.}\label{fig:calibcooling}
\end{figure}

The calibration discussed below was done with the data described in Table \ref{table:modelinout} obtained from 03/08/2017 to 30/08/2017. As it can be seen in Figure \ref{fig:paretofrontier_cooling}, the NSGA-II solver has found not only one optimal set of RC parameters but a list of Pareto solutions. The best compromise solution, $\hat{\sigma}$, is the one in red whose power absolute error ($f_2(\hat{\sigma}$)) equals 150,628$kW$ with relative error ($P_{\textrm{error}}$) of 5.06\% and whose MAPE value for indoor temperature ($f_1(\hat{\sigma}$)) is 1.25\% with median deviation of 0.23$\degree$C. This solution is situated in the ``knee'' of the approximate Pareto frontier, where a small change in its position causes a large change in at least one of the objectives. 

The thermal power and indoor temperature curves associated to the best compromise solution are shown in Figure \ref{fig:tempplotscooling} and Figure \ref{fig:powerplotscooling}, respectively. As it can be seen, the forecast curves are relatively close to the actual ones, which proves that the model not only captures well the dynamics of both indoor temperature and power consumption, but it is also precise regarding the magnitude of the actual values using only 4 weeks of historical data.

\subsubsection{Optimization strategy results}
\label{subsec:coolinggopt}

In the cooling strategy, there are two decision variables in principle:
\begin{enumerate}
    \item Day initial time: switch over from night set-point to comfort (day) set-point;
    \item Night initial time: switch over from day set-point to night set-point.
\end{enumerate}
After carrying out tests with optimization, we noticed that the starting time proposed by BatIntel was not different from the one already used by the BatIntel building and that the savings obtained by the optimization was solely due to the night initial time. This shows that the anticipated cooling of the building has no expected positive impact on its daily power consumption. For this reason, the work done on the cooling optimization strategy was focused only on the second variable, the night initial time. This also vindicates our choice for setting $t_{beg}^{savings}=7PM$ and $t_{end}^{savings}=10PM$ in the calculation of the energy savings.

Table \ref{table:optsavcooling} shows the estimated absolute and relative energy savings obtained by BatIntel over a 10-day period while Table \ref{table:opterrorcooling} provides the error of the model for the same period. In summary, the BatIntel solution allowed to obtain an average savings of 8.6\%, i.e 1362.3$kWh$. It is important to notice, however, that no “expert” audit has been done in order to estimate the maximum possible savings for the BatIntel building. Therefore, it is not possible to evaluate whether the savings obtained by the BatIntel solution was small or large when compared to the maximum possible savings. 

\begin{table}[ht]
\caption{Synthesis of the cooling optimization strategy results from 10 days between August and September 2017 with the estimated savings obtained.}
\centering
\begin{tabular}{c  c  c  c}
 \hline
 \textbf{Day} & \textbf{Date} & \textbf{Energy savings in $kWh$} & \textbf{Energy savings in \%} \\
 \hline
 1  & 09/08/2017 & 971  & 6.6 \\
 2  & 10/08/2017 & 1235 & 10.2 \\
 3  & 11/08/2017 & 1640 & 10.9 \\
 4  & 14/08/2017 & 2033 & 11.1 \\
 5  & 23/08/2017 & 542  & 2.5 \\
 6  & 30/08/2017 & 906  & 3.6 \\
 7  & 04/09/2017 & 1418 & 9.5 \\
 8  & 05/09/2017 & 1952 & 10.2 \\
 9  & 06/09/2017 & 1418 & 9.9 \\
 10 & 07/09/2017 & 1508 & 11.6 \\
 \hline
    & \textbf{Mean} & \textbf{1362.3} & \textbf{8.6} \\
    & \textbf{S.D.} & \textbf{441.9} & \textbf{3.1} \\
\hline
\end{tabular}
\label{table:optsavcooling}
\end{table}

\begin{table}[ht]
\caption{Synthesis of the cooling optimization strategy results from 10 days between August and September 2017 with the accuracy of the model.}
\centering
\begin{tabular}{c  c  c  c  c}
 \hline
 \textbf{Day} & \textbf{Date} & \textbf{Power error in $kW$} & \textbf{Power error in \%} & \textbf{Indoor temp. error in \%} \\
 \hline
 1  & 09/08/2017 & 38.75 & 11 & 2.6 \\
 2  & 10/08/2017 & 61.25 & 21 & 1.9 \\
 3  & 11/08/2017 & 37.75 & 11 & 2.3 \\
 4  & 14/08/2017 & 42.25 & 9  & 2.8 \\
 5  & 23/08/2017 & 37.75 & 10 & 3 \\
 6  & 30/08/2017 & 37.25 & 8  & 2.9 \\
 7  & 04/09/2017 & 43.75 & 13 & 0.9 \\
 8  & 05/09/2017 & 37.75 & 11 & 3.1 \\
 9  & 06/09/2017 & 45.5 & 12 & 2.6 \\
 10 & 07/09/2017 & 41 & 16 & 2.4 \\
 \hline
    & \textbf{Mean} & \textbf{42.3} & \textbf{12.2} & \textbf{2.4} \\
    & \textbf{S.D.} & \textbf{6.8}  & \textbf{3.6}  & \textbf{0.6} \\
\hline
\end{tabular}
\label{table:opterrorcooling}
\end{table}

The errors shown in Table \ref{table:opterrorcooling} are calculated from the definitions of $T_{\textrm{error}}$ and $P_{\textrm{error}}$ in \eqref{eq:temppowererrors}, where the optimal decision variables have been used for obtaining both the forecast and the actual values. The power error is about 12.2\% in average while the indoor temperature error is of 2.4\% in average. Notice that the uncertainty of the model is partly ascribed to the uncertainty inherent in the weather forecast. For instance, the mean errors in external temperature and in solar irradiance over the period are of 7\% and 11\%, respectively. Although these values are not too large, they still make an impact on the performance of the BatIntel model as the uncertainties are propagated. More statistics on the weather forecast data are given in Table \ref{table:weatherforecastcooling}. 

\begin{table}[ht]
\caption{Statistics on the weather forecast data used as input to the model during the cooling season.}
\centering
\begin{tabular}{l c c}
 \hline
 \textbf{Description} & \textbf{External temperature} & \textbf{Solar irradiance} \\
 \hline
 Mean relative error & 7\% & 11\% \\
 Mean absolute error & 1.4$\degree$C & 80$W/m^2$\\
 Maximum absolute error & 6$\degree$C & 600$W/m^2$\\
\hline
\end{tabular}
\label{table:weatherforecastcooling}
\end{table}

\begin{figure}
    \centering
    \begin{subfigure}[t]{0.5\textwidth}
        \centering
        \includegraphics[height=3in]{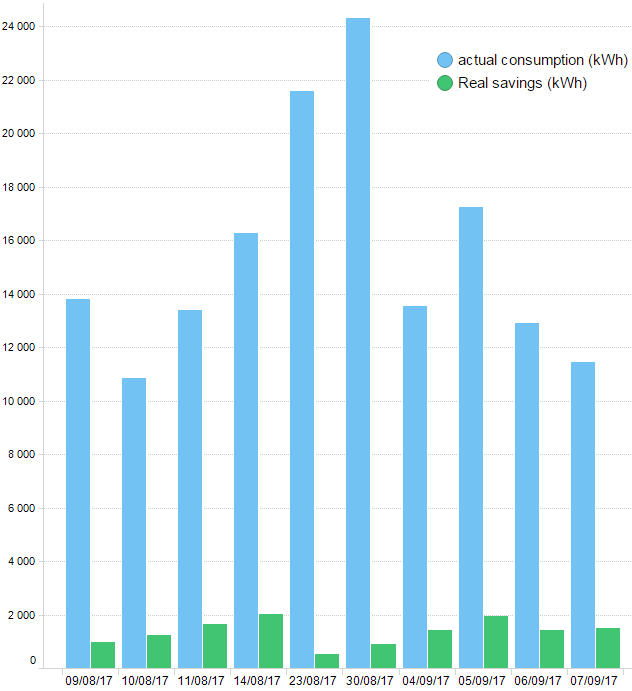}
        \caption{In blue, the actual energy consumption of the building, and, in green, the energy savings obtained by BatIntel.}
        \label{fig:gain_cooling}
    \end{subfigure}%
    ~ 
    \begin{subfigure}[t]{0.5\textwidth}
       \centering
        \includegraphics[height=3in]{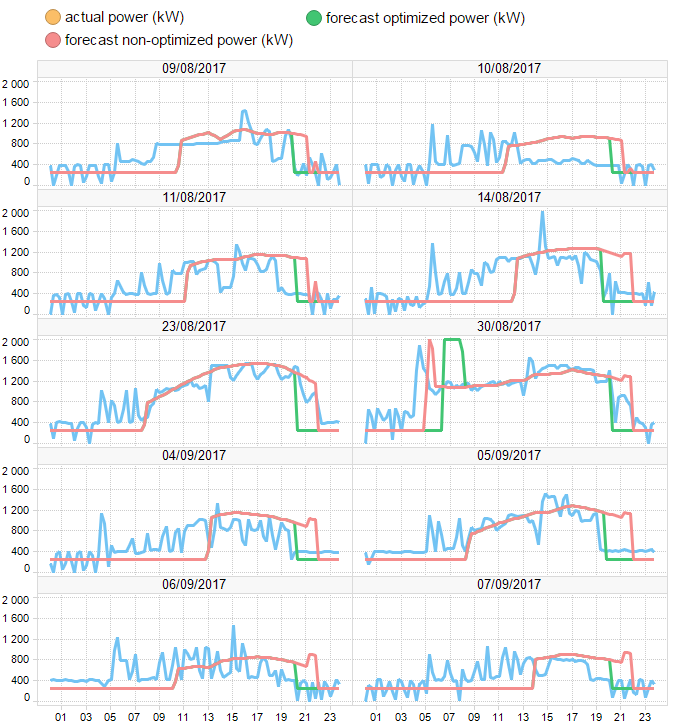}
        \caption{Hourly optimization results for each day. In blue, the actual power; in green, the forecast optimized power; and, in red, the forecast non-optimized power.}
        \label{fig:optimized_days_cooling}
    \end{subfigure}
    \caption{Optimization strategy results from the cooling season with data from 09/08/2017 to 07/09/2017.}\label{fig:optcooling}
\end{figure}

The Figure \ref{fig:gain_cooling} illustrate in detail the savings obtained by BatIntel as well as the actual energy consumption of each day. Moreover, the Figure \ref{fig:optimized_days_cooling} shows the actual power, the forecast non-optimized power and the forecast optimized power for each optimized day of the cooling season. Notice that the actual power is the result of the application of the BatIntel recommended actions based on the forecast optimized power given by the model and on the optimization strategy results; therefore, both curves (actual power and forecast optimized power) are typically close to each other at the night initial time --- the only optimization variable --- value proposed by BatIntel, but not necessarily in the remaining hours as there is no other recommended action to be taken.

\subsection{Heating season}
\label{subsec:heatingseason}

\subsubsection{Calibration results}
\label{subsec:heatingcalib}

\begin{figure}
    \centering
    \begin{subfigure}[t]{\textwidth}
        \centering
        \includegraphics[height=2.5in]{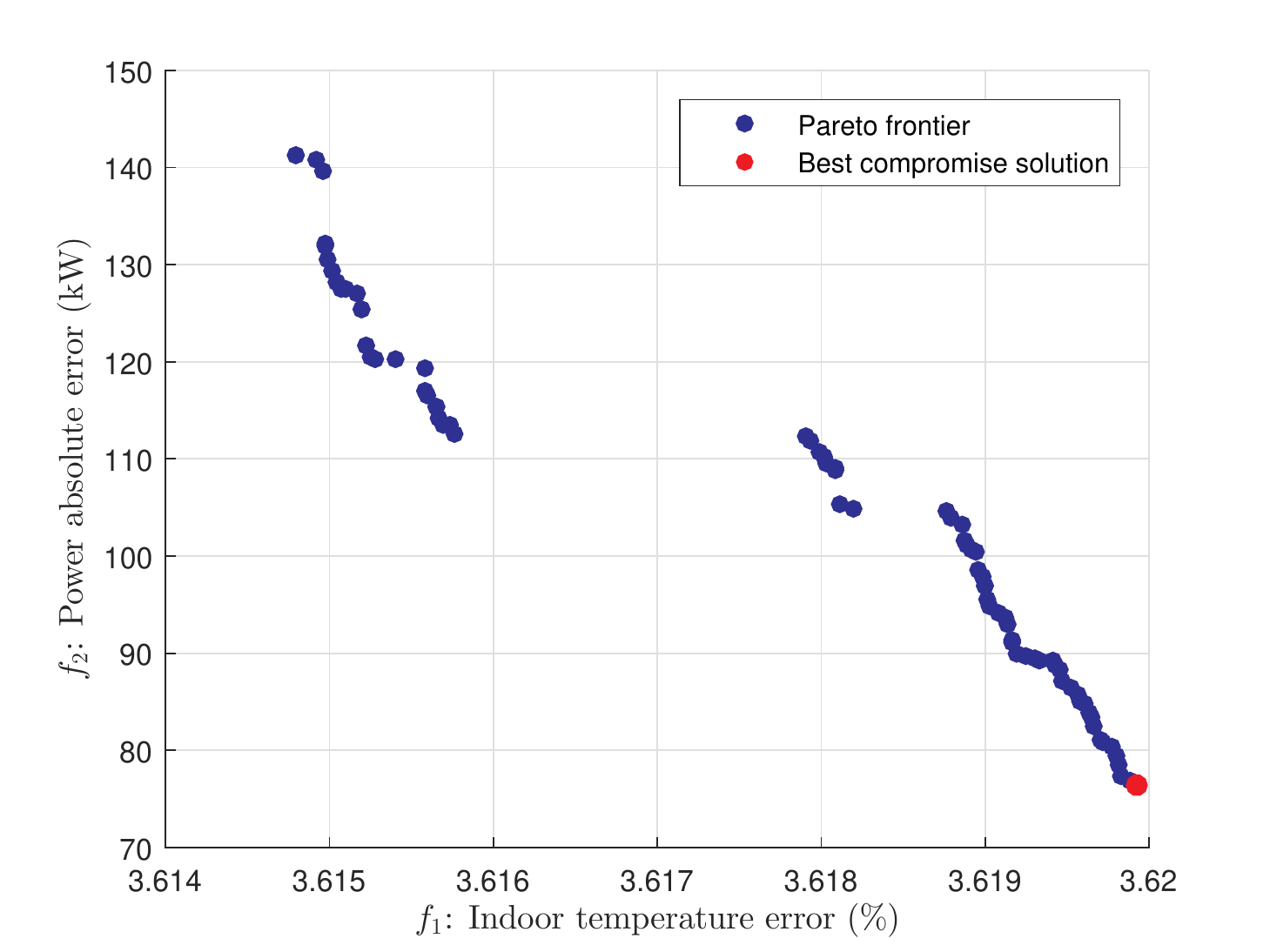}
        \caption{Pareto frontier obtained from the heating season. In the vertical axis, the median absolute deviation in power ($kW$), and, in the horizontal axis, the relative error in indoor temperature.}
        \label{fig:paretofrontier_heating}
    \end{subfigure}
    ~ 
    \begin{subfigure}[t]{0.5\textwidth}
        \centering
        \includegraphics[height=2in]{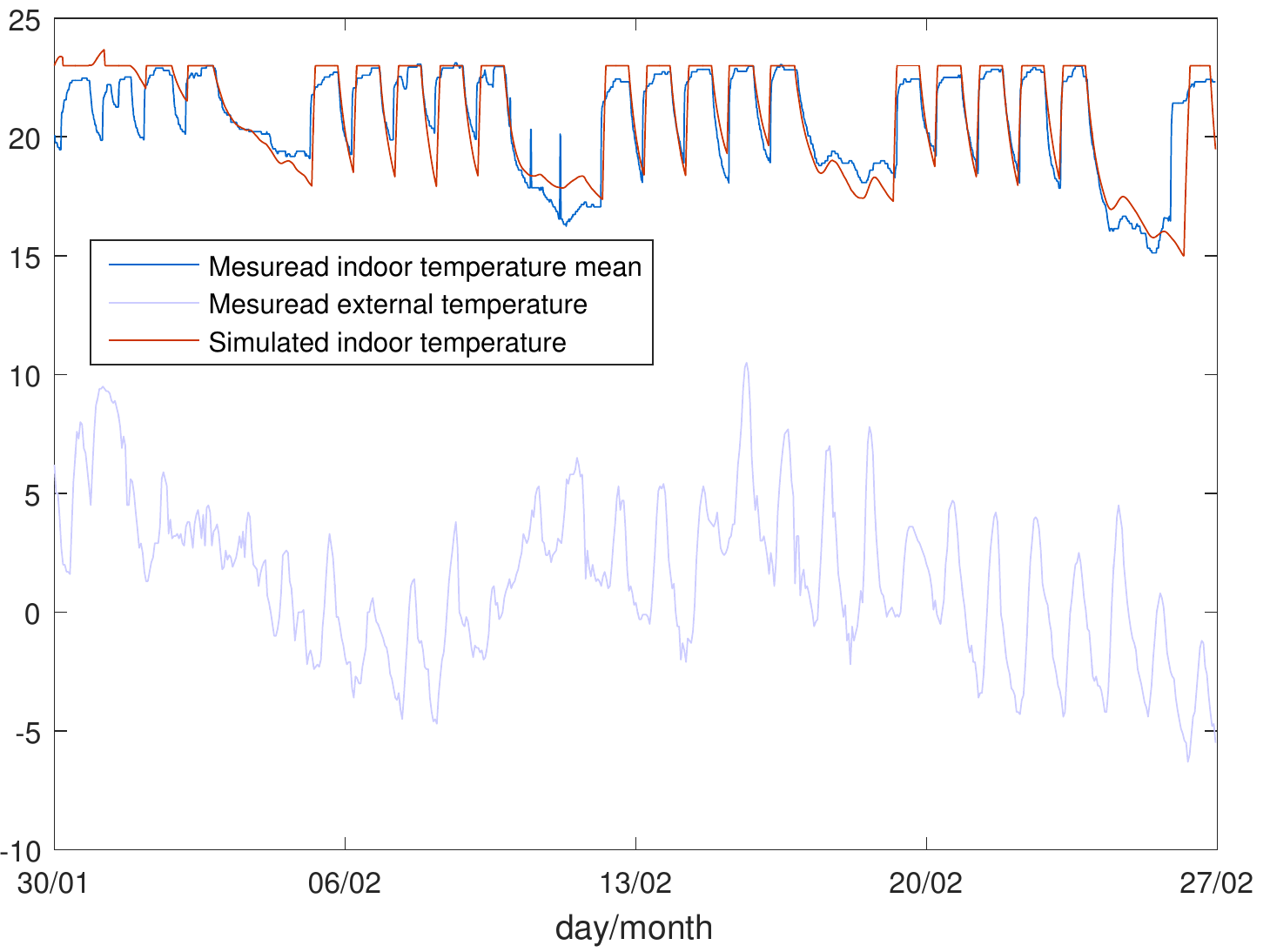}
        \caption{Temperature curves ($\degree$C) obtained with the best compromise solution from the Pareto frontier over the heating season. In blue, actual indoor temperatures; in red, indoor temperatures predicted by the model; and, in violet, actual external temperatures.}
        \label{fig:tempplotsheating}
    \end{subfigure}%
    ~ 
    \begin{subfigure}[t]{0.5\textwidth}
       \centering
        \includegraphics[height=2in]{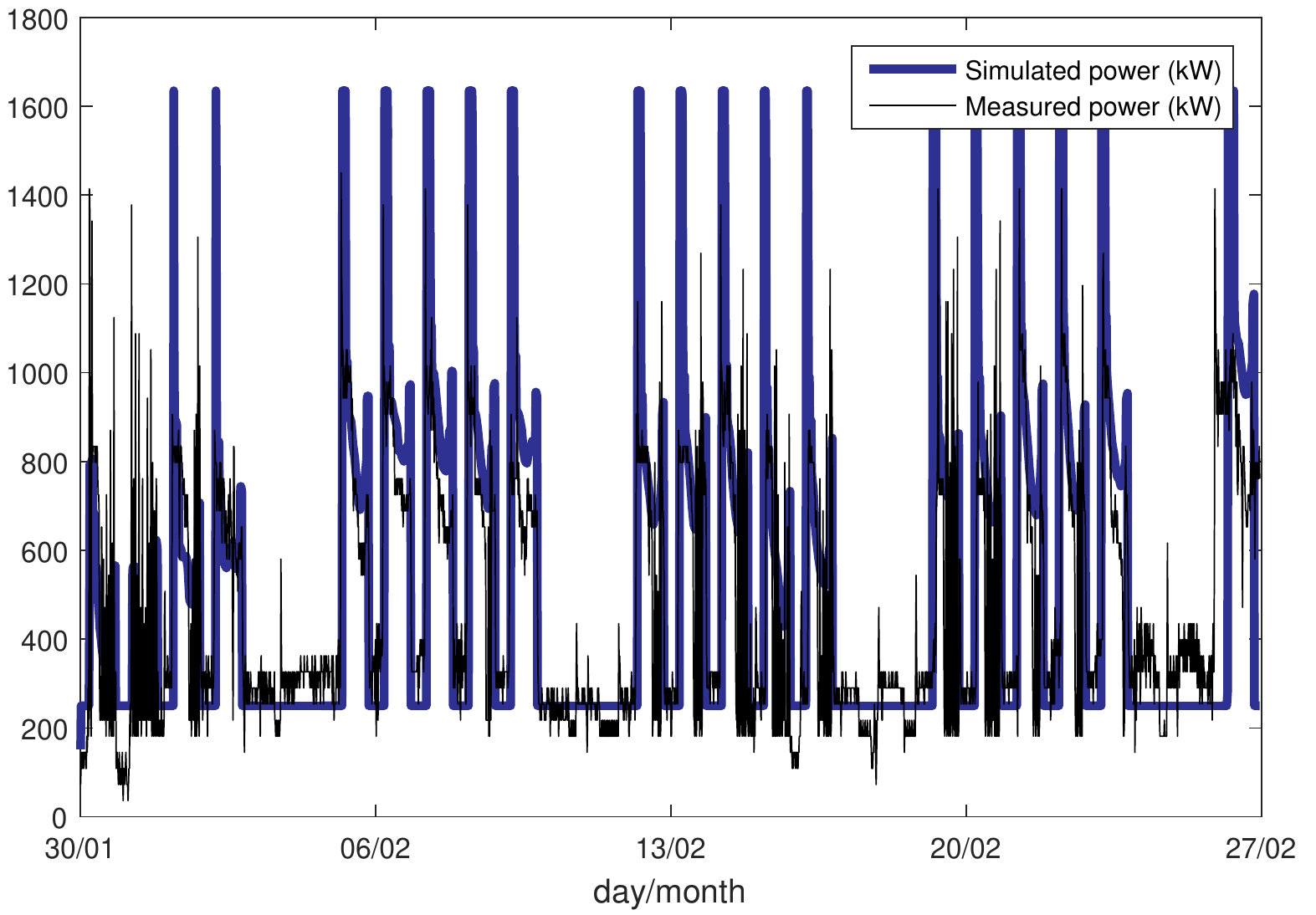}
        \caption{Power curves obtained with the best compromise solution from the Pareto frontier over the heating season. In black, the actual power, and, in blue, the power predicted by the model.}
        \label{fig:powerplotsheating}
    \end{subfigure}
    \caption{Calibration results from the heating season with data from 31/01/18 to 27/02/18.}\label{fig:calibheating}
\end{figure}

The calibration was done with the data obtained from 31/01/18 to 27/02/18. The Figure \ref{fig:paretofrontier_heating} shows the approximate Pareto frontier obtained from NSGA-II. The best compromise solution is situated at the lower right corner of the graph with power absolute error of 87.34$kW$ and power relative error of 6.2\%, and MAPE value for indoor temperature of 3.62\% with median deviation of 0.5$\degree$C. Since the variance of the temperature error is quite low among Pareto solutions, the best compromise solution is located in the lower right part of the Pareto boundary to favor a lower power error.

The thermal power and indoor temperature curves associated to the best compromise solution are shown in Figure \ref{fig:tempplotsheating} and Figure \ref{fig:powerplotsheating}, respectively. We can conclude that the model has clearly identified the dynamics of the actual indoor temperature and power curves and that the curve fitting is successful, although there is a slight overestimation in power during some days. As in the case of cooling, we observe that the gap in indoor temperature is very small whereas the one in power consumption is larger, probably due to the fact that there is more variance in power than in indoor temperature and that the signal of the power is quite noisy.

\subsubsection{Optimization strategy results}
\label{subsec:heatingopt}

The BatIntel optimization heating strategy was applied on 17 days between 15/02/18 and 06/04/19, providing an average savings of 11.7\% (1198.2$kWh$). This value is higher than the one of 8.6\% obtained in the cooling season, which shows that our approach is more likely to give better results when heating systems are optimized rather than cooling machines for this type of building. However, we must consider the fact that all 3 decisions variables described in Section \ref{subsec:strategydesc} were optimized in the heating case, while only the night initial time was optimized in the cooling demonstration. Besides, the time interval for the calculation of the savings is given by $t_{beg}^{savings}=4AM$ and $t_{end}^{savings}=8PM$, which is much wider than in the cooling case. Therefore, there are more possibilities of improvement in the heating strategy.

The model average errors are of 8.2\% in power and of 2.4\% in indoor temperature. A more detailed analysis of the savings obtained over this period are given in Table \ref{table:optsavheating}, while in Table \ref{table:opterrorheating} we present the model errors in a daily basis. We also give some statistics on the weather forecast data used in the heating season in Table \ref{table:weatherforecastheating}.

\begin{table}[htb]
\caption{Synthesis of the heating optimization strategy results from 17 days between February and April 2018 with the estimated savings obtained.}
\centering
\begin{tabular}{c  c  c  c}
 \hline
 \textbf{Day} & \textbf{Date} & \textbf{Energy savings in $kWh$} & \textbf{Energy savings in \%} \\
 \hline
 1  & 15/02/2018 & 2310 & 20.1 \\
 2  & 16/02/2018 & 1738 & 14.5 \\
 3  & 20/02/2018 & 2248 & 17.3 \\
 4  & 21/02/2018 & 1533 & 13.0 \\
 5  & 22/02/2018 & 1129 & 9.7 \\
 6  & 23/02/2018 & 397  & 3.3 \\
 7  & 28/02/2018 & 636  & 4.4 \\
 8  & 20/03/2018 & 1797 & 15.3 \\
 9  & 21/03/2018 & 714  & 6.0 \\
 10 & 22/03/2018 & 1062 & 10.6 \\
 11 & 23/03/2018 & 1290 & 13.1 \\
 12 & 27/03/2018 & 384  & 5.2 \\
 13 & 28/03/2018 & 1725 & 19.6 \\
 14 & 29/03/2018 & 813  & 8.4 \\
 15 & 30/03/2018 & 1179 & 16.5 \\
 16 & 05/04/2018 & 40   & 0.6 \\
 17 & 06/04/2018 & 1375 & 20.3 \\
 \hline
    & \textbf{Mean} & \textbf{1198.2} & \textbf{11.7} \\
    & \textbf{S.D.} & \textbf{631.8}  & \textbf{6} \\
\hline
\end{tabular}
\label{table:optsavheating}
\end{table}

\begin{table}[htb]
\caption{Synthesis of the heating optimization strategy results from 17 days between February and April 2018 with the accuracy of the model.}
\centering
\begin{tabular}{c  c  c  c  c}
 \hline
 \textbf{Day} & \textbf{Date} & \textbf{Power error in $kW$} & \textbf{Power error in \%} & \textbf{Indoor temp. error in \%} \\
 \hline
 1  & 15/02/2018 & 221.7 & 15.6 & 4.0 \\
 2  & 16/02/2018 & 114.3 & 7.9  & 2.7 \\
 3  & 20/02/2018 & 154.6 & 11.0 & 2.9 \\
 4  & 21/02/2018 & 89 & 7.2  & 1.3 \\
 5  & 22/02/2018 & 125.4 & 10.1 & 2.1 \\
 6  & 23/02/2018 & 89 & 6.4  & 2.5 \\
 7  & 28/02/2018 & 143.2 & 12.0 & 4.5 \\
 8  & 20/03/2018 & 138.1 & 9.6  & 3.2 \\
 9  & 21/03/2018 & 67.1 & 5.2  & 3.2 \\
 10 & 22/03/2018 & 56.7 & 4.6  & 1.1 \\
 11 & 23/03/2018 & 37.7 & 2.7  & 1.3 \\
 12 & 27/03/2018 & 89 & 12.2 & 0.8 \\
 13 & 28/03/2018 & 72.5 & 5.3  & 1.5 \\
 14 & 29/03/2018 & 93 & 12.3 & 1.1 \\
 15 & 30/03/2018 & 56.8 & 4.2  & 0.8 \\
 16 & 05/04/2018 & 63.7 & 5.6  & 3.9 \\
 17 & 06/04/2018 & 57.6 & 7.2  & 3.4 \\
 \hline
    & \textbf{Mean} & \textbf{98.2} & \textbf{8.2} & \textbf{2.4} \\
    & \textbf{S.D.} & \textbf{45.3}  & \textbf{3.5}  & \textbf{1.2} \\
\hline
\end{tabular}
\label{table:opterrorheating}
\end{table}

\begin{table}[htb]
\caption{Statistics on the weather forecast data used as input to the model during the heating season.}
\centering
\begin{tabular}{l c c}
 \hline
 \textbf{Description} & \textbf{External temperature} & \textbf{Solar irradiance} \\
 \hline
 Mean relative error & 4.5\% & 9.6\% \\
 Mean absolute error & 0.4$\degree$C & 11$W/m^2$\\
 Maximum absolute error & 2.8$\degree$C & 302$W/m^2$\\
\hline
\end{tabular}
\label{table:weatherforecastheating}
\end{table}

\begin{figure}
    \centering
    \begin{subfigure}[t]{0.5\textwidth}
        \centering
        \includegraphics[height=2.5in]{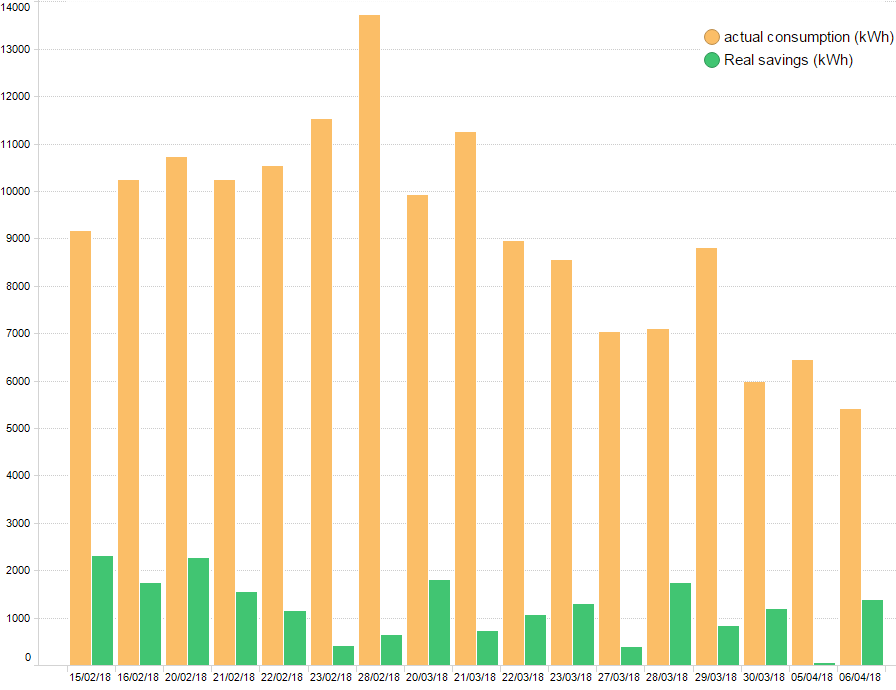}
        \caption{In blue, the actual energy consumption of the building, and, in green, the energy savings obtained by BatIntel.}
        \label{fig:gain_heating}
    \end{subfigure}%
    ~ 
    \begin{subfigure}[t]{0.5\textwidth}
       \centering
        \includegraphics[height=2.5in]{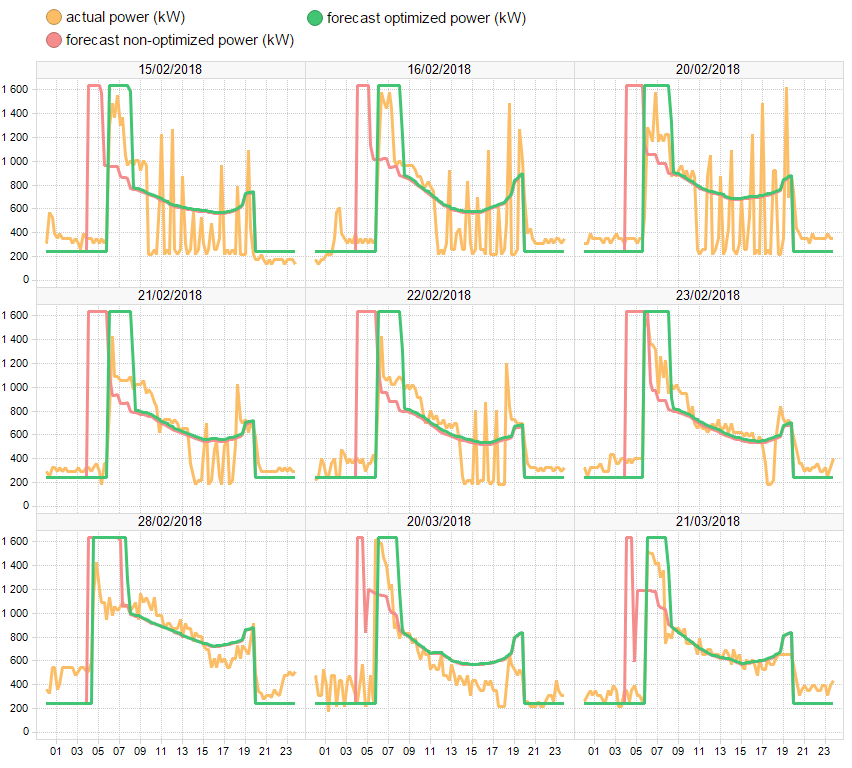}
        \caption{Daily optimization results from 15/02/2018 to 21/03/18. In blue, the actual power; in green, the forecast optimized power; and, in red, the forecast non-optimized power.}
        \label{fig:optimized_days_heating1}
    \end{subfigure}
    ~
    \begin{subfigure}[t]{\textwidth}
       \centering
        \includegraphics[height=2.5in]{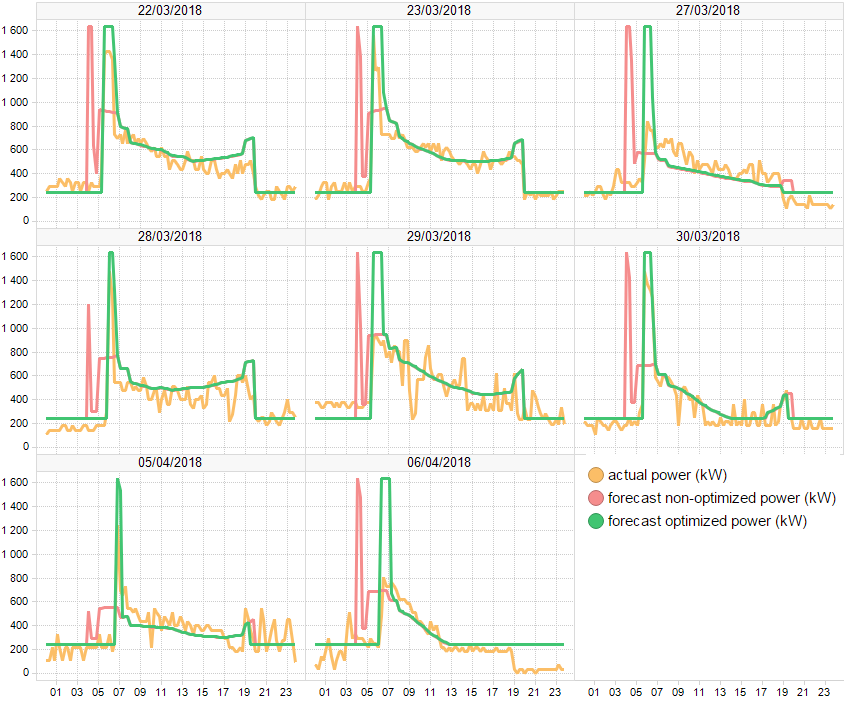}
        \caption{Daily optimization results from 22/03/2018 to 06/04/18.}
        \label{fig:optimized_days_heating2}
    \end{subfigure}
    \caption{Optimization strategy results from the heating season with data from 15/02/2018 to 06/04/18.}\label{fig:optheating}
\end{figure}

\clearpage

\section{Conclusions}
\label{sec:conclusions}

We have presented a model-based predictive control approach that optimizes the planning of heating and cooling systems for tertiary sector buildings. It consists of a grey-box model that predicts the building heat load and indoor temperature and of optimization strategies that make use of this model in order to reduce the total energy consumption of the next day. The proposed approach was able to yield an energy reduction of 8.6\% (1362.3$kWh$) during the cooling season and of 11.7\% (1198.2$kWh$) during the heating season in a case study of a 25,000m$^2$ commercial building containing three gas boilers and two refrigeration units. 

Based on the experimental results shown in this paper, we conclude by stating two key points: (1) the confirmed ability of the BatIntel model to represent the thermal behaviour of monozone buildings and to forecast their power consumption and indoor temperature based on a short period of historical data; and (2) the efficiency of the optimization strategy for reducing the energy consumption of this type of building while satisfying thermal comfort requirements.

We have identified possible improvements for the current approach as the next steps in order to increase its performance and to make it applicable to a larger extent of buildings. For instance, the extension of a R6C2-type model to handle different thermal zones in a building is possible and useful when the building has different zones being heated and cooled in parallel. This extension would allow us to address buildings where the thermal behaviour is more complex without having to adapt the optimization strategy, but only the building model.

Besides optimizing the night set-point temperature, one could also optimize the day set-point temperature (i.e. the set-point indoor temperature to be attained in the building during day time) and try to obtain an even lower energy consumption. In a higher level of control, a day set-point temperature for each working hour during the day could be considered in the optimization strategy.

\section{Acknowledgment}
\label{sec:acknowledgement}

We thank our colleagues from Veolia Belgium for their assistance during the course of this research and we would also like to show our gratitude to the partner that allowed this research work to be conducted with one of their buildings.

\bibliographystyle{unsrt}
\bibliography{references}

\end{document}